\documentclass{elsart}
\journal{Annals of Physics}

\usepackage{epsfig}

\usepackage{amsmath}
\usepackage{amssymb}

\begin{document}
\numberwithin{equation}{section}

\begin{frontmatter}


\title{Measurement-induced decoherence and Gaussian smoothing of the
Wigner distribution function}
\author{Yong-Jin Chun} and
\author{Hai-Woong Lee}
\ead{hwlee@laputa.kaist.ac.kr}
\address{Department of Physics, Korea Advanced Institute of
Science and Technology, Daejeon, 305-701, Korea}

\begin{abstract}

We study the problem of measurement-induced decoherence using the
phase-space
approach employing the Gaussian-smoothed Wigner distribution function. Our
investigation is based on the notion that measurement-induced decoherence is
represented by the transition from the Wigner distribution to the
Gaussian-smoothed Wigner distribution with the widths of the smoothing
function identified as measurement errors. We also compare the smoothed
Wigner distribution with the corresponding distribution resulting from the
classical analysis. The distributions we computed are the phase-space
distributions for simple one-dimensional dynamical systems such as a
particle
in a square-well potential and a particle moving under the influence of a
step potential, and the time-frequency distributions for high-harmonic
radiation emitted from an atom irradiated by short, intense laser pulses.
\end{abstract}

\begin{keyword}
measurement \sep decoherence \sep Gaussian smoothing \sep Wigner
distribution function \sep high harmonic generation
\PACS 03.65.Ta \sep 03.65.Yz \sep 42.65.Ky
\end{keyword}
\end{frontmatter}

\section{Introduction}
Quantum phase-space distribution functions \cite{PR106_121,PR259_147}
offer an alternative means of
formulating quantum mechanics to the standard wave mechanics formulation.
Some of the properties that these functions possess, however, make it
difficult to associate them with a direct probabilistic interpretation. The
Wigner distribution function \cite{PR40_749}, for example, can take on
negative values, while
the Husimi distribution function \cite{PP22_264},
although nonnegative, does not yield the
correct marginal distributions. These functions are therefore usually
regarded as a mathematical tool used to describe the quantum behavior of the
system being considered and are thus referred to as the quasiprobability
distribution functions.
\\  \indent
There, however, have been continued theoretical investigations
\cite{BST44_725,PRL52_1064,AP110_102} on the
possibility of the physical significance of the Husimi distribution function
and more generally of the nonnegative smoothed Wigner distribution function,
ever since Arthurs and Kelly \cite{BST44_725}
showed that the Husimi distribution is a proper
probability distribution associated with a particular model of simultaneous
measurements of position and momentum. An operational definition of a
probability distribution can be given
\cite{PRL52_1064} that explicitly takes into account the
action of the measurement device modeled as a ``filter". The analysis shows
that the phase-space distribution connected to a realistic simultaneous
measurement of position and momentum can be expressed as a convolution of the
Wigner function of the system being detected and the Wigner function of the
filter state. If the Wigner function of the filter state is given by a
minimum uncertainty Gaussian function, which is the case for an ideal
simultaneous measurement with the maximal accuracy allowed by the Heisenberg
uncertainty principle, the distribution can be identified as the Husimi
distribution. A generalization to the case of a measurement with less
accuracy leads immediately to an identification of the nonnegative smoothed
Wigner distribution as a distribution connected to a realistic measurement.
\\  \indent
It is then clear that, in the phase-space approach, the act of a measurement
is conveniently modeled by Gaussian smoothing, with the widths of the
smoothing function identified as measurement errors. The phase-space approach
based on the smoothed Wigner distribution function thus provides a convenient
framework in which to study any changes that occur in the phase-space
distribution of a system as a result of a measurement.
\\  \indent
In this paper we aim to investigate effects of a measurement on the
phase-space distribution of a system using the
phase-space description based on the
smoothed Wigner distribution function \cite{PH83A_210}.
We consider some simple one-dimensional
systems---a particle in a square-well potential, a particle moving under the
influence of a step potential, and a one-dimensional model atom
irradiated by high-power laser pulses emitting high-harmonic radiation---and
compute their Wigner and smoothed Wigner distribution functions. The pure
quantum distribution of the systems unaffected by a measurement is
represented by contour curves of the Wigner distribution functions, whereas
the ``coarse-grained" distribution, {\it i.e.}, the distribution
``contaminated" by a measurement is displayed by contour curves of the smoothed
Wigner distribution functions. It may be expected that delicate quantum
features are obscured more strongly by a measurement with larger measurement
errors. Such effects of a measurement, which may be referred to as
measurement-induced decoherence, can be studied by observing changes in the
contour curves of the smoothed Wigner distribution functions, as the widths
of the smoothing Gaussian function are increased. We also compare the contour
curves of the smoothed Wigner distributions we computed with the
corresponding contour curves resulting from the classical analysis. This
should yield information on what quantum features survive measurement-induced
decoherence.
\\  \indent
In Section \ref{WSW} we briefly describe the Wigner and
smoothed Wigner distribution
functions and discuss their physical significance in relation to a
measurement. In Section \ref{ODS} we choose,
as our examples of simple dynamical
systems, a particle in an infinite square-well potential and a particle
subjected to a step potential and compute their contour curves of the Wigner
and smoothed Wigner distribution functions. These contour curves form the
basis of our discussion of measurement-induced decoherence. Another example
we take for our study is a one-dimensional model atom irradiated by
high-power laser pulses, which is described in Section \ref{HHG}.
The time-frequency
Wigner and smoothed Wigner distributions are computed for high-harmonic
radiation emitted from the model atom, and the effect of a measurement
on the time-frequency distribution is discussed.
Finally a discussion is given in Section \ref{DIS}.
\section{Wigner and smoothed Wigner distribution functions}\label{WSW}
The Wigner distribution function \cite{PR40_749} is defined by
\begin{equation}
W(q,p) = \frac{1}{\pi \hbar} \int
dx e ^{-2ipx/\hbar} \Psi^\ast (q-x)
\Psi(q+x),
\end{equation}
where $q$ and $p$ are the coordinate and the momentum of the system being
considered and $\Psi$ is the wave function. In our study of one-dimensional
dynamical systems, we restrict our attention to the simple case when the
system is in its energy eigenstate. The wave function $\Psi$ and the Wigner
distribution function $W$ is then independent of time. The Gaussian-smoothed
Wigner distribution function \cite{PH83A_210} is given by
\begin{equation}
G(q,p) = \frac{1}{2\pi \sigma_q \sigma_p} \int dq^ \prime  \int dp^\prime
e ^{-\frac{(q^\prime - q)^2}{2 \sigma_q ^2}}
e ^{-\frac{(p^\prime - p)^2}{2 \sigma_p ^2}} W(q^\prime, p^\prime).
\label{GSW}
\end{equation}
When the smoothing Gaussian function is a minimum uncertainty wave packet,
{\it i.e.}, when $\sigma_q \sigma_p = \frac{\hbar}{2}$, $G$ can be identified
as the Husimi distribution function,
\begin{equation}
H(q,p) = \frac{1}{2\pi \hbar}\sqrt{\frac{\kappa}{\pi \hbar}}
\left| \int dx e ^{-\kappa (x-q)^2/2\hbar }
e ^{-ipx/\hbar} \Psi(x)\right|^2  ,
\end{equation}
where the parameter $\kappa$ defined as $\kappa=\frac{\hbar}{2\sigma_q^2}=
\frac{2\sigma_p^2}{\hbar}$ has the dimension of $mass/time$. If $G$ is to
represent a probability distribution corresponding to a simultaneous
measurement of $q$ and $p$, the widths should satisfy the inequality $\sigma_
q \sigma_p \geq \frac{\hbar}{2}$, where $\sigma_q$ and $\sigma_p$,
respectively, can be identified as measurement errors in $q$ and $p$ of the
detection apparatus. For general mathematical treatments, however, one may
also include the unphysical regime $\sigma_q \sigma_p < \frac{\hbar}{2}$
in the analysis . In particular, in the limit $\sigma_q \to 0$ and $\sigma_p
\to 0$, $G(q,p)$ approaches the Wigner function $W(q,p)$.
\\  \indent
In general the Wigner function can be defined in space of any pair of
conjugate variables. In particular, in studies of signal processing, the
Wigner function in the time-frequency space, $W(t,\omega)$,
plays an important role \cite{IEEE77_941}.
The main tool for our time-frequency analysis of high-harmonic
radiation generated by an atom irradiated by laser pulses, which is described
in Section \ref{HHG}, is the Wigner time-frequency distribution function of the
dipole acceleration
\begin{equation}
W(t,\omega) = \frac{1}{\pi} \int_{-\infty} ^{\infty}
d\tau e ^{-2i\omega\tau} \ddot d ^\ast (t-\tau)
\ddot d(t+\tau), \label{TimeWigner}
\end{equation}
where the dipole acceleration $\ddot d (t)$ is given by
\begin{equation}
\ddot d(t) = \frac{d^2}{dt^2} d(t)=\langle \Psi(t) \left| \ddot x \right|
\Psi(t) \rangle. \label{da}
\end{equation}
In Eq. (\ref{da}),
$\Psi (t)$ represents the wave function of the electron in the atom
and $x$ is the position of the electron. The Gaussian-smoothed Wigner
distribution function in time-frequency space is given by
\begin{equation}
G(t,\omega) = \frac{1}{2\pi \sigma_t \sigma_\omega} \int dt^ \prime
\int d\omega^\prime e ^{-\frac{(t^\prime - t)^2}{2 \sigma_t ^2}}
e ^{-\frac{(\omega^\prime - \omega)^2}{2 \sigma_\omega ^2}}
W(t^\prime, \omega^\prime). \label{TimeGSW}
\end{equation}
When the widths $\sigma_t$ and $\sigma_\omega$ satisfy $\sigma_t
\sigma_\omega =\frac{1}{2}$, $G$ becomes the Husimi time-frequency
distribution function
\begin{equation}
H(t,\omega) = \frac{1}{2\pi}\sqrt{\frac{\kappa}{\pi}}
\left| \int d\tau e ^{-\kappa (\tau-t)^2/2 }
e ^{-i\omega \tau} \ddot d(\tau)\right|^2  ,\label{TimeHusimi}
\end{equation}
where the parameter $\kappa$ given by $\kappa=\frac{1}{2\sigma_t ^2}=2\sigma_
\omega ^2$ has now the dimension of $(time)^{-2}$.
\\  \indent
Since the measurement errors satisfying the inequality $\sigma_
q \sigma_p \geq \frac{\hbar}{2}$ (or $\sigma_
t \sigma_\omega \geq \frac{1}{2}$) belong to the physical regime, the contour
curves of $G$ associated with the smoothing Gaussian function satisfying this
inequality can in principle be observed. Perhaps the most straightforward way
of constructing such contour curves from measurements is to make a large
number of simultaneous measurements of position and momentum (or time and
frequency) upon identically prepared systems. Each measurement should be
performed with the same measurement errors $\sigma_q$ and $\sigma_p$ (or
$\sigma_t$ and $\sigma_\omega$). Each measurement should also be performed on
a given system no more than once, because the measurement disturbs the system
and its phase-space distribution. From the results of such measurements
one can determine the phase-space distribution (or the time-frequency
distribution) for the
system under consideration. The contour curves of the smoothed Wigner
distribution function associated with the widths $\sigma_q$ and $\sigma_p$
(or $\sigma_t$ and $\sigma_\omega$) of the smoothing Gaussian function are
then obtained by connecting the phase-space points (or the time-frequency
points) having the same phase-space probability (or the same time-frequency
space probability).
\\  \indent
Our main concern here is the effects of a measurement on quantum properties
of a system. Information on such effects can be directly obtained by
comparing contour curves of the Wigner distribution function and those of
various smoothed Wigner distribution functions associated with different
values of $\sigma_q$ and $\sigma_p$ (or $\sigma_t$ and $\sigma_\omega$). In
the next two sections we present results for our computation of the contour
curves in phase space for two simple one-dimensional dynamical systems and in
time-frequency space for high-harmonic radiation emitted by an atom
irradiated by laser pulses.
\section{One-dimensional dynamical systems in a stationary state}\label{ODS}
In this section we present results of our computation of contour curves of
the Wigner and smoothed Wigner distribution functions
for two simple systems in a
stationary state; a particle in a symmetric infinite square-well potential
and a particle moving under the influence of a step potential. For stationary
systems it is possible to obtain contour plots of the Wigner distribution
function directly by solving a pair of time-independent equations of motion
satisfied by the Wigner function \cite{PCP60_581}.
One can also derive a pair of time-independent
equations of motion for the smoothed Wigner distribution function
and can thus obtain its contour plots from the solutions of these equations.
In the present work, however, we have chosen to numerically integrate the
Wigner distribution function according to Eq. (\ref{GSW})
in order to obtain contour
curves of the smoothed Wigner distribution function, because an analytical
form for the Wigner distribution function is known for each of the two
systems being considered here.
\subsection{Particle in a square well}
We consider a particle in a symmetric infinite square-well potential
\begin{equation}
V(q) =
\begin{cases}
0, &\text{if $-a \le q \le a$,} \\
\infty, &\text{if $q < -a$, or $q > a$}.
\end{cases}
\end{equation}
An analytic form for the Wigner distribution function is known for this
system in its eigenstate \cite{FP13_61}.
For all our computation we take the mass of the
particle to be 1 and the width of the well to be 20 in an arbitrary unit
system in which $\hbar=1$.
\\  \indent
Shown in Figs. \ref{fig1}(a) and \ref{fig1}(b)
are contour plots of the Wigner distribution function and the
smoothed Wigner distribution function, respectively, for the
particle in its ground state. The
smoothing parameters for Fig. \ref{fig1}(b) are chosen to be
$\sigma_q = 3.62$ and $\sigma_
p = 0.157$ ($\sigma_q \sigma_p=0.57 > \frac{\hbar}{2}$). Comparing
Figs. \ref{fig1}(a) and \ref{fig1}(b),
we see that the smoothed Wigner distribution function has a simpler
structure than the nonsmoothed Wigner distribution function. This reflects
the effect of measurement-induced decoherence. The most conspicuous quantum
structure which is exhibited in Fig. \ref{fig1}(a)
but is not revealed when observed with
$\sigma_q = 3.62$ and $\sigma_p = 0.157$ is closed orbits off the origin
called ``islands". Two pairs of the islands are shown in
Fig. \ref{fig1}(a). Close inspection of the contour plot of the
Wigner distribution function, however,
reveals many such islands, although not shown in the figure. We note that a
purely classical dynamical consideration would indicate that the contour
lines of the classical probability follow classical trajectories. Thus, for
the present case of a particle in a square-well potential, classical contour
lines would consist of a set of pairs of straight horizontal lines in the
region $-a\le q \le a$. It is interesting to note that the contour curves of
the smoothed Wigner function, Fig. \ref{fig1}(b), are not straight lines,
which indicate
that some quantum characteristics still survive Gaussian smoothing. The
quantum dynamical property represented by curved contour curves may be
referred to as nonlocality, because these curves lead one to interpret that
the particle changes its momentum even before it actually hits the potential
wall. (For classical dynamical interpretation of Wigner contour curves and
problems arising from it, see \cite{FP13_61} and \cite{FP22_995}.)
This quantum behavior (curved contour
curves) lies within the limits of observation because
Fig. \ref{fig1}(b) belongs to the
physical regime ($\sigma_q \sigma_p \geq \frac{\hbar}{2}$).
\\  \indent
Figs. \ref{fig2}(a)-\ref{fig2}(d) show contour curves for the
same particle in its fifth eigenstate \cite{JPA23_1025}. The
contour plot of the Wigner distribution function shown
in Fig. \ref{fig2}(a) indicates
that there are five maxima along the $q$ axis arising from the five-peak
standing wave structure of the eigenfunction and two maxima along the $p$
axis corresponding to the momenta to the right and left, respectively, of the
classical particle of the same energy. Characteristic quantum features such
as curved contour lines and the formation of islands are even more evident
here. Figs. \ref{fig2}(b) and \ref{fig2}(c) are contour plots of the Husimi
distribution function ($
\sigma_q \sigma_p = \frac{\hbar}{2}$) with $\sigma_q=0.637$, $\sigma_p=0.785$,
and with $\sigma_q=5.70$, $\sigma_p=0.0877$, respectively.  One sees that a
choice of a large $\sigma_q$ ($\sigma_p$) results in strong smoothing along
the $q$ ($p$) axis. These two figures represent the quantum phase-space
distribution connected to ideal simultaneous measurements with different
degrees of the position {\it vs}. momentum uncertainties. The fact that
Husimi curves are different for a different choice of $\sigma_q$ or $\sigma_p
$ means physically that the quantum distribution looks different depending on
how one observes it. It should be noted that the curves of
Figs. \ref{fig2}(b) and \ref{fig2}(c),
although simpler in structure than the curves of Fig. \ref{fig2}(a)
due to measurement-induced decoherence,
still exhibit strong quantum behavior such as the
islands even though they belong to the physical regime. This can be
understood if we note that the uncertainties $\Delta_q \equiv \sqrt{\langle q
^2 \rangle - \langle q \rangle^2}$ and $\Delta_p \equiv \sqrt{\langle p^2
\rangle - \langle p \rangle^2}$ associated with the fifth eigenstate are
given by $\Delta_q=5.70$ and $\Delta_p=0.785$ ($\Delta_q \Delta_p=4.48$). The
widths $\sigma_q$ and $\sigma_p$ of the smoothing Gaussian function used for
Figs. \ref{fig2}(b) and \ref{fig2}(c) are less than the uncertainties
$\Delta_q$ and $\Delta_p$ inherent
in the fifth eigenstate, and therefore the probabilistic nature of the
quantum eigenstate is expected to be displayed by the corresponding contour
curves. Fig. \ref{fig2}(d) shows contour plots for the case
$\sigma_q=\Delta_q$ and $\sigma_p=\Delta_p$.
One sees now that the islands, a clear indication of a
strong quantum feature, have disappeared. Note, however, that the nonlocal
nature ({\it i.e.}, curved contour curves) is still indicated by the contour
curves of Fig. \ref{fig2}(d), as these curves deviate
from the corresponding classical
trajectories, {\it i.e.}, straight horizontal lines.
\subsection{Particle incident on a potential step}
As a second example we consider a particle of fixed energy $E$ moving under
the influence of a step potential
\begin{equation}
V(q) =
\begin{cases}
0, &\text{if $q <0$,} \\
V_0, &\text{if $q \ge0$.}
\end{cases}
\end{equation}
The wave function and the Wigner distribution function for this particle
have been given earlier \cite{FP13_61}.
The parameter values we have chosen for our computation
are $m=1$, $V_0 =1$, $E=0.5$ in an arbitrary unit system in which $\hbar =1$.
Since the particle energy $E$ is one half the potential step $V_0$, the wave
function decreases exponentially with respect to $q$ in the region $q\geq 0$.
\\  \indent
Fig. \ref{fig3}(a) shows a contour plot of the Wigner distribution
function for the
particle. In addition to the quantum features such as curved contour lines
and the islands that are already exhibited by the Wigner curves of the
particle in a square-well potential, we see that some curves exhibit
tunneling into the potential step.
\\  \indent
In Figs. \ref{fig3}(b) and \ref{fig3}(c) we show contour plots of the
Husimi distribution function for
the cases $\sigma_q = 0.5$, $\sigma_p =1$ and $\sigma_q=1.58$, $\sigma_p =
0.316$, respectively. Strong quantum features such as nonlocality, islands
and tunneling are still clearly shown, although the curves belong to the
physical regime. This can be understood by recalling that, for the particle
of energy $E=0.5$ being considered here, $\Delta_q \to \infty$ (since the
particle is extended in the entire half space $q < 0$) and $\Delta_p =
\sqrt{2 m E}= 1$, and the widths $\sigma_q$ and $\sigma_p$ chosen for
Figs. \ref{fig3}(b) and \ref{fig3}(c)
are still sufficiently small that the quantum nature of the eigenfunction
survives.
\\  \indent
Finally in Fig. \ref{fig3}(d) we show a contour plot of the smoothed
Wigner distribution
function with $\sigma_q=2.236$, $\sigma_p=1$. Although the islands have
disappeared now, the curves still exhibit nonlocality and tunneling. The
curves are not quite the same as the classical trajectories which consist of
a set of pairs of straight lines in the region $q < 0$ for $E < V_0$.

\section{High-harmonic radiation generated by a one-dimensional atom
irradiated by laser pulses}\label{HHG}
The phenomenon of high-harmonic generation in atomic gases irradiated by
high-power laser pulses has been investigated intensively in the past both
theoretically and experimentally \cite{RPP60_389}.
A typical emission spectrum observed
experimentally shows a broad plateau extending to high-order harmonics
accompanied with a fast cutoff. The mechanism by which high-order harmonics
are generated is now well understood. In particular, the simple classical
three-step model \cite{PRL68_3535} has been quite useful in providing
the conceptual understanding of this phenomenon.
\\  \indent
In this work we focus on the time-frequency distribution of the emitted
radiation, {\it i.e.}, on the relationship between the time of emission and
the frequency of emitted radiation. Since the temporal variation of the
emitted signal can be represented by the dipole acceleration
\cite{PRA51_1458}, the pure quantum-mechanical time-frequency
distribution of the emitted radiation is described
by the Wigner time-frequency distribution of the dipole acceleration as
defined in Eq. (\ref{TimeWigner}) \cite{PRA63_063403}.
Our main interest lies in the effect of measurement on this
Wigner time-frequency distribution. In this particular case, the measurement
should consist of the simultaneous measurement of time and frequency. One
needs to perform a large number of such measurement upon a large number of
identically prepared samples (atomic gases irradiated by laser pulses). The
time-frequency distribution associated with the measurements with the
uncertainties $\sigma_t$ and $\sigma_\omega$ is given by the
Gaussian-smoothed Wigner distribution $G(t,\omega)$ of Eq. (\ref{TimeGSW}).
We also compare this
smoothed Wigner distribution with the corresponding distribution resulting
from classical analysis.
\\  \indent
In subsection \ref{SYS} we give a brief description of the system.
In subsection \ref{CLASS} we
then find the classical time-frequency distribution based on the three-step
model. The main part of this section is subsection \ref{QUANT}
in which quantum time-frequency distribution, both Wigner and
smoothed Wigner, are presented.

\subsection{System}\label{SYS}
The system under consideration is a one-dimensional model atom irradiated by
laser pulses. The wave function of the electron in the atom evolves with time
according to the time-dependent Schr\"odinger equation which reads in atomic
units
\begin{equation}
i \frac{\partial}{\partial t} | \Psi(t) \rangle =
\left[ \frac{p^2}{2}+V(x) - x E(t) \right] | \Psi(t) \rangle ,
\label{SE}
\end{equation}
where the atomic potential $V(x)$ is modeled upon a soft-core potential of
the form
\begin{equation}
V(x)= - \frac{1}{\sqrt{\beta^2 + x^2}} .
\end{equation}
We set $\beta \cong 0.67$ a.u. in order for the system to have a ground-state
energy equivalent to the binding energy $I_p$ of Ne.
The laser pulse incident on
the atom is assumed to be a Gaussian pulse with the electric field $E(t)$
given by
\begin{equation}
E(t) = E_0 \exp \left[-2 \ln 2 \frac{t^2}{(\Delta t)^2}
\right] \cos (\omega_L t) .
\end{equation}
In our calculations, we take the center frequency of the pulse, $\omega_L =
800$ nm, the full width at half maximum of the laser pulse,
$\Delta t =160$ fs, and the peak intensity $3 \times 10^{14}$ $\rm W/cm^2$.

\subsection{Classical analysis}\label{CLASS}
Here we consider the question of when each harmonic is emitted. If one uses
classical analysis based on the three-step model \cite{PRL68_3535},
it is straightforward to obtain the relation between the time of
emission and the frequency of emitted
radiation. After tunneling through the Coulomb barrier, different electrons
in atoms at different locations in general see the laser field at different
phases and therefore follow different classical trajectories. Consequently,
different electrons return to their nuclei at different times with different
kinetic energies. Let us say that the $i$th electron returns to its nucleus
at time $t_i$ with kinetic energy $K_i = \frac{1}{2}m v_i^2$. When this
electron recombines with the nucleus, the radiation of frequency $\omega_i$
given by $\hbar \omega_i = I_p + K_i$ is emitted. Collecting $\omega_i$ and $
t_i$ for all different electrons, one has the answer to the question of when
the radiation of a given frequency is emitted. The solid curves in
Fig. \ref{fig4} shows
the result of such classical calculations. The curves exhibit the
characteristic $\Lambda$ structure, which indicates that, within one-half
optical cycle of the laser field, there exist two different classical
trajectories, called ``short" and ``long" paths, with two different return
times that lead to an emission of radiation of a given frequency
\cite{PRL74_3776}. The open
circles in Fig. \ref{fig4} represent contributions from multiple
recollisions. There is
a possibility that recombination does not occur when the electron returns to
its nucleus. One thus needs to consider the cases when the electron
recombines with the nucleus at the $n$th ($n \geq 2$) encounter
with the nucleus.
These open circles together with the solid curves of the $\Lambda$ structure
comprise the classical time-frequency distribution obtained using the
three-step model.

\subsection{Quantum-mechanical analysis}\label{QUANT}
We now turn to a quantum time-frequency distribution of the emitted
radiation \cite{PRA63_063403}. For our computation of the
Wigner distribution of Eq. (\ref{TimeWigner}), we first solve the time
dependent Schr\"odinger equation, Eq. (\ref{SE}),
using the Crank-Nicolson method.
Once $\Psi (t)$ is known, the dipole acceleration $\ddot d (t)$ can be
calculated using Eq. (\ref{da})
and then the Wigner time-frequency distribution $W(t,
\omega)$ can be obtained using Eq. (\ref{TimeWigner}).
The Gaussian-smoothed Wigner
distribution $G(t,\omega)$ can then be obtained using
Eq. (\ref{TimeGSW}). In particular, the Husimi time-frequency
distribution can be computed relatively easily by using
Eq. (\ref{TimeHusimi}).
\\  \indent
The Wigner time-frequency distribution $W(t, \omega)$ we computed shows a
very complicated structure, and it is difficult to extract physical meaning
out of it. We therefore choose not to show it here and just to mention that it
consists of a large number of small islands, a clear indication that strong
quantum behavior is exhibited by the time-frequency distribution.
\\  \indent
In Figs. \ref{fig5}(a)-\ref{fig5}(c), we show Gaussian-smoothed
Wigner time-frequency distributions we computed. The widths of the
smoothing Gaussian function are $\sigma_t = 2.236$ and
$\sigma_\omega =0.224$ for Fig. \ref{fig5}(a). Since $\sigma_t
\sigma_\omega=\frac{1}{2}$, Fig \ref{fig5}(a) is the Husimi
time-frequency distribution. For Figs. \ref{fig5}(b) and
\ref{fig5}(c), we have $\sigma_t=2.236$, $\sigma_\omega=0.448$,
and $\sigma_t=4.472$, $\sigma_\omega=0.224$, respectively. We note
that the overall shape of the smoothed Wigner distributions
suggests the $\Lambda$ structure of the classical distribution of
Fig. \ref{fig4}. The similarity, however, is only qualitative,
because the smoothed Wigner distributions have many detailed
structures (many islands) absent in the classical distribution. We
mention, however, that the smoothed distributions of Figs.
\ref{fig5}(a)-\ref{fig5}(c) are much simpler than the nonsmoothed
Wigner distribution not shown. We see from Figs.
\ref{fig5}(a)-\ref{fig5}(c) that the distribution in the low-order
region takes on the maximum value at times 0.5, 1.0, 1.5, 2.0,
{\it etc}, {\it i.e.} at times at which the laser field amplitude
takes on the maximum or minimum value. The $\Lambda$-like
structure in the midplateau region shown in Figs.
\ref{fig5}(a)-\ref{fig5}(c) indicates that photons in the plateau
region are emitted mainly at two different times within one-half
cycle of the laser field, in agreement with the classical analysis
based on Fig. \ref{fig4}. The separation between these two times
decreases and eventually vanishes as the harmonic order is
increased toward the cutoff. In the cutoff region, therefore,
photons are emitted once within one-half cycle. Figs.
\ref{fig5}(a)-\ref{fig5}(c) show that the distribution in the
cutoff region is maximum at times 0.7, 1.2, 1.7, 2.2, {\it etc},
indicating that photons in the cutoff region are emitted at times
just before the laser amplitude vanishes.

\section{Discussion} \label{DIS}
In this paper we have studied effects of a measurement on the
phase-space (time-frequency) distribution using the approach based
on the smoothed Wigner distribution function. The physical
significance of the phase-space (time-frequency) contour curves
differs depending on the widths $\sigma_q$ and $\sigma_p$
($\sigma_t$ and $\sigma_\omega$) of the smoothed Wigner
distribution function used. While the contour curves of the
nonsmoothed Wigner distribution function with $\sigma_q \to 0$ and
$\sigma_p \to 0$ ($\sigma_t \to 0$ and $\sigma_\omega \to 0$)
represent the unobservable pure quantum distribution, the Husimi
curves with $\sigma_q \sigma_p = \frac{\hbar }{2}$ ($\sigma_t
\sigma_\omega = \frac{1}{2}$) represent the quantum distribution
observed in ideal simultaneous measurements with the maximal
accuracy allowed by the Heisenberg uncertainty principle. In
general, contour curves in the physical regime corresponding to
$\sigma_q \sigma_p \geq \frac{ \hbar}{2}$ ($\sigma_t \sigma_\omega
\geq \frac{1}{2}$) represent the quantum distribution associated
with a coarse-grained observation caused by measurement errors. We
have seen that quantum features such as the islands tend to
disappear and the resulting quantum distribution becomes simpler
in structure, as the measurement errors are increased. In general,
however, the smoothed Wigner distributions associated with
relatively large measurement errors have more structures than the
corresponding classical distributions.
\\  \indent
In conclusion we have shown, with simple one-dimensional model systems as
examples, that the phase-space formulation based on the smoothed Wigner
distribution function provides a very natural framework in which to study
how the quantum phase-space distribution is affected by a measurement.
\ack
This research was supported by Korea Research Foundation
under Contract No. 2001-015-DP0107.



\begin{figure}
\begin{center}
\includegraphics[width=1.0\columnwidth]{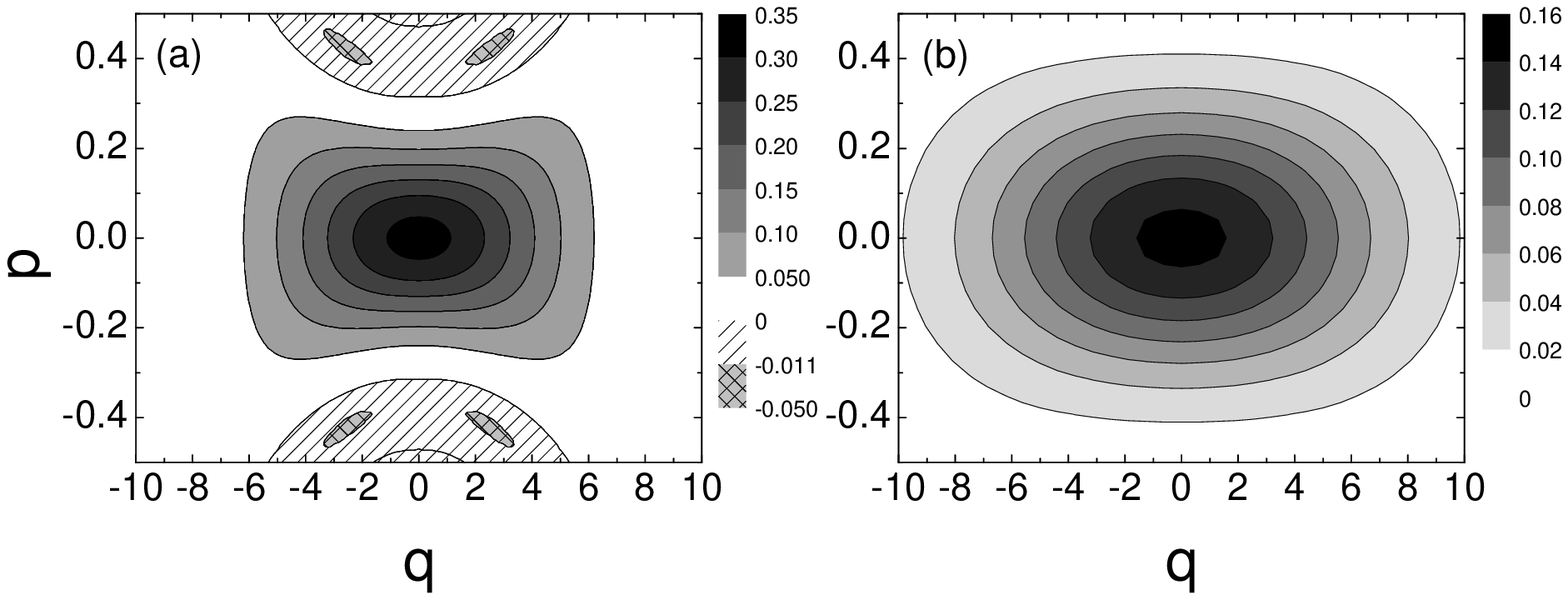}
\vspace{0.3cm} 
\caption{
\label{fig1}}
\end{center}
\end{figure}

\begin{figure}
\begin{center}
\includegraphics[width=1.0\columnwidth]{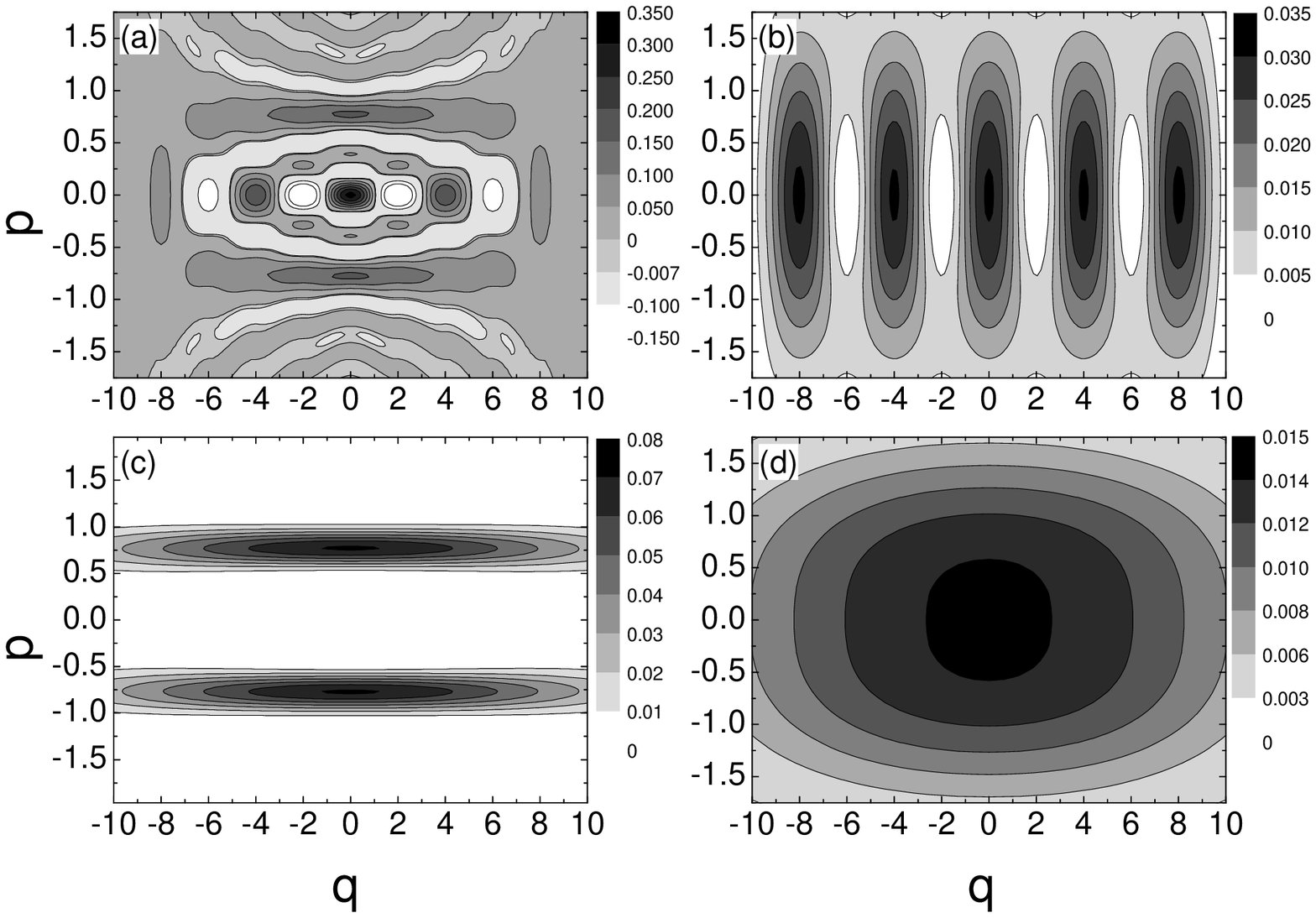}
\vspace{0.3cm} 
\caption{
\label{fig2}}
\end{center}
\end{figure}

\begin{figure}
\begin{center}
\includegraphics[width=1.0\columnwidth]{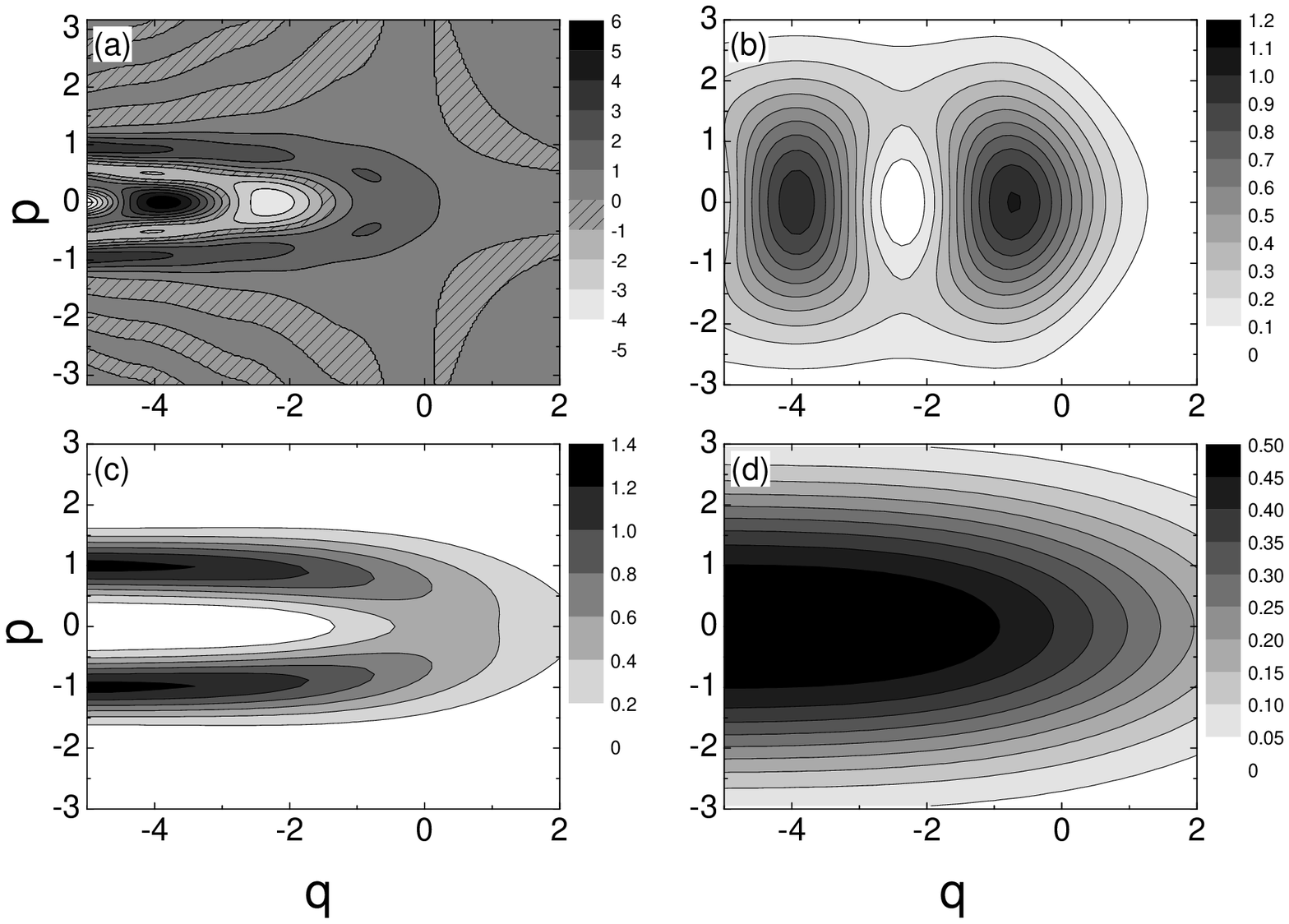}
\vspace{0.3cm} 
\caption{
\label{fig3}}
\end{center}
\end{figure}

\begin{figure}
\begin{center}
\includegraphics[width=1.0\columnwidth]{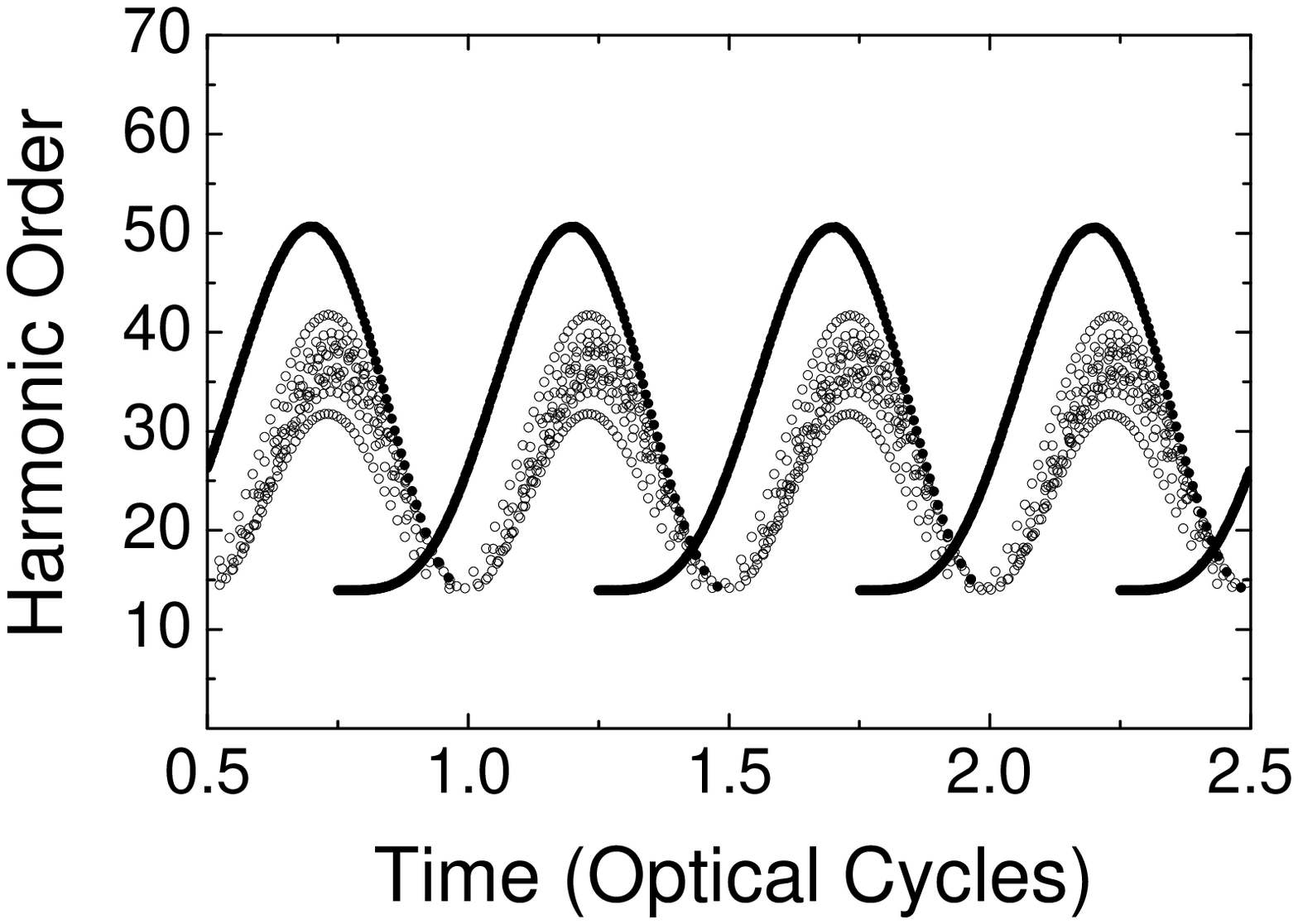}
\vspace{0.3cm} 
\caption{
\label{fig4}}
\end{center}
\end{figure}

\begin{figure}
\begin{center}
\includegraphics[width=.8\columnwidth]{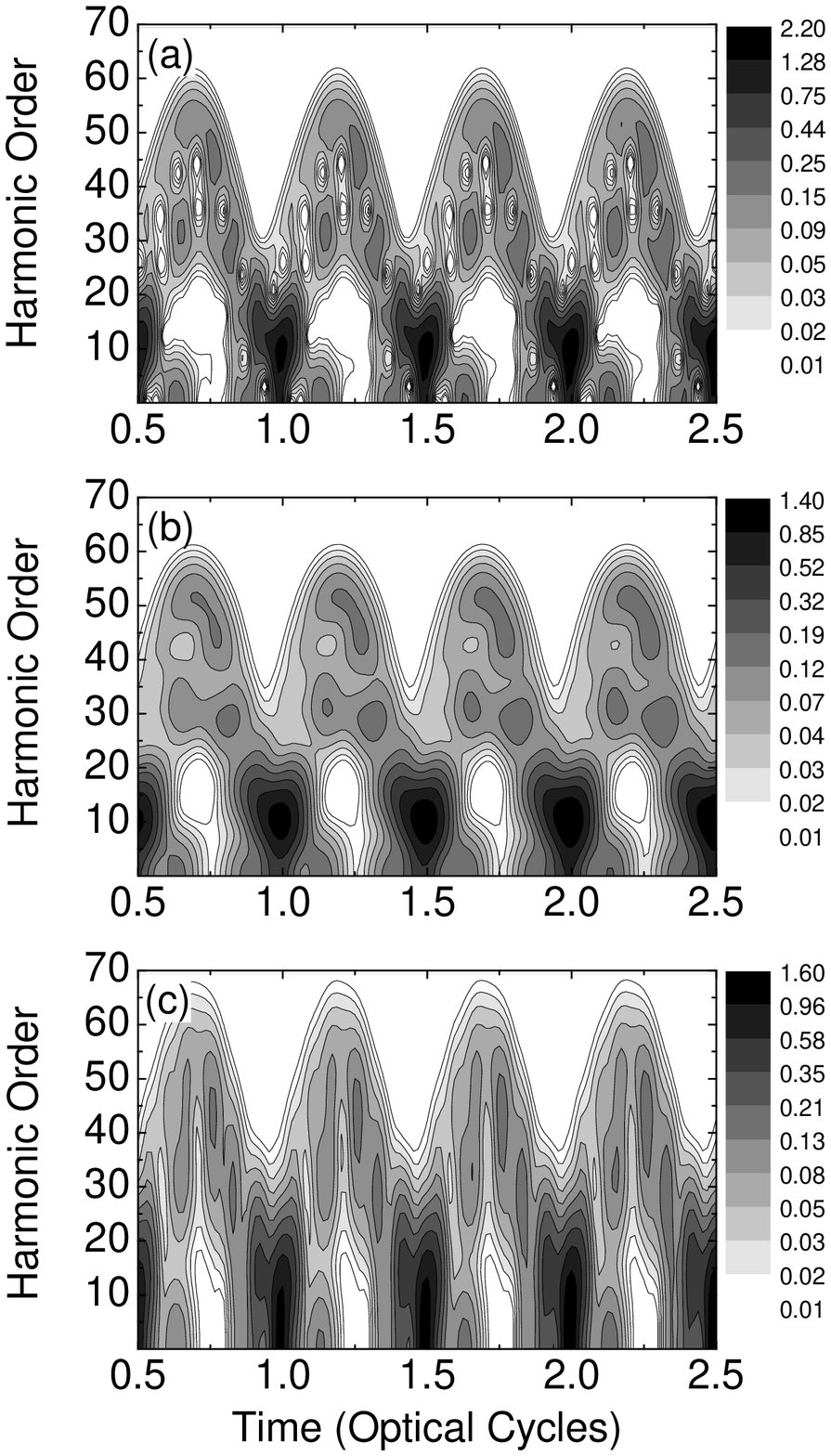}
\vspace{0.3cm} 
\caption{
\label{fig5}}
\end{center}
\end{figure}

\newpage
\begin{flushleft}
{\Large \bf Figure Captions}
\vspace{\baselineskip}

{\bf Figure \ref{fig1}.} Contour plots for a particle in a symmetric
infinite square-well potential in its ground state. The mass of the particle
is 1 and the width of the potential well is 20 in an arbitrary unit
system in which $\hbar=1$. (a) Wigner function, (b) smoothed Wigner
function with $\sigma_q =3.62$ and $\sigma_p = 0.157$.

{\bf Figure \ref{fig2}.} Same as Figure \ref{fig1} except that the particle
is in its fifth eigenstate. (a) Wigner function,
(b) Husimi function with $\sigma_q =0.637$, $\sigma_p = 0.785$,
(c) Husimi function with $\sigma_q =5.70$, $\sigma_p = 0.0877$,
(d) smoothed Wigner function with $\sigma_q =5.70$, $\sigma_p = 0.785$.

{\bf Figure \ref{fig3}.} Contour plots for a particle moving under the
influence of a step potential. The mass and energy of the particle are
1 and 0.5, respectively, and the value of the potential is 0 for
$q<0$ and 1 for $q\geq0$, in an arbitrary unit system in which
$\hbar=1$.
(a) Wigner function,
(b) Husimi function with $\sigma_q=0.5$, $\sigma_p=1$,
(c) Husimi function with $\sigma_q=1.58$, $\sigma_p=0.316$,
(d) smoothed Wigner function with $\sigma_q=2.236$, $\sigma_p=1$.

{\bf Figure \ref{fig4}.} Classical time-frequency distribution for
high-harmonic radiation. The solid curves represent contributions from the
short and long paths, and the open circles represent contributions from
multiple recollisions.

{\bf Figure \ref{fig5}.} Contour plots for high-harmonic radiation.
(a) Husimi function with $\sigma_t=2.236$, $\sigma_\omega=0.224$,
(b) smoothed Wigner function with $\sigma_t=2.236$, $\sigma_\omega=0.448$,
(c) smoothed Wigner function with $\sigma_t=4.472$, $\sigma_\omega=0.224$.
\end{flushleft}
\newpage

\clearpage

\end{document}